\documentclass[aps,twocolumn,superscriptaddress,a4paper]{revtex4}
\usepackage{amsfonts,amssymb,amsmath}
\usepackage[utf8]{inputenc}
\usepackage[pdftex,breaklinks=true,colorlinks=true]{hyperref}
\usepackage[pdftex]{graphicx}
\usepackage{leftidx}

\newcommand{\ket}[1]{\left|#1\right>}

\newcommand{\mean}[1]{\langle#1\rangle}
\begin{document}

\title{Inverse participation ratios in the XXZ spin chain}

\author{Grégoire Misguich, Vincent Pasquier and Jean-Marc Luck}
\affiliation{Institut de Physique Th\'eorique, Universit\'e Paris Saclay, CEA, CNRS, F-91191 Gif-sur-Yvette, France}

\date{\today}

\begin{abstract}
We investigate numerically the inverse participation ratios in a spin-1/2 XXZ chain,
computed in the ``Ising'' basis ({\it i.e.,} eigenstates of $\sigma^z_i$).
We consider in particular a quantity $T$, defined by summing the inverse participation ratios of all the eigenstates in the zero magnetization sector
of a finite chain of length $N$, with open boundary conditions. From a dynamical point of view,
$T$ is proportional to the stationary return probability to an initial basis state, averaged over all the basis states (initial conditions).
We find that $T$ exhibits an exponential growth, $T\sim\exp(aN)$, in the gapped phase of the model and a linear scaling,
$T\sim N$, in the gapless phase. These two different behaviors are analyzed in terms of the distribution of the participation ratios of individual eigenstates.
We also investigate the effect of next-nearest-neighbor interactions, which break the integrability of the model.
Although the massive phase of the non-integrable model also has $T\sim\exp(aN)$, in the gapless phase $T$ appears to saturate to a constant value.
\end{abstract}
\maketitle

\section{Introduction}

In recent years there has been a renewed interest in the dynamical properties of out-of-equilibrium isolated quantum systems~\cite{Polkovnikov_RMP,goldstein_long-time_2010,eisert_quantum_2015}.
For instance, in a ``quantum quench'' setup, the system is initially prepared in the ground state of some Hamiltonian $\mathcal{H}_0$
and then evolved (unitarily) with a Hamiltonian $\mathcal{H}\ne \mathcal{H}_0$.
A fruitful point of view is to consider a small spatial region of a large system and to investigate the long-time limit of observables defined in this subsystem.
Some typical questions one may then address are whether these expectation values have well-defined long-time limits and if they are described by some statistical ensemble (either thermal or not)~\cite{ETH2008}.
Such problems still represent an active subject of research.
Clearly,
the steady states that may be reached crucially depend on the structure of the eigenstates of $\mathcal{H}$.
In particular, it has been shown that integrable systems generally fail to reach thermal states~\cite{Rigol_GGE2007}. Somewhat analogously, disordered many-body systems
in the ``many-body localized phase'' also fail to ``thermalize,'' a phenomenon which has attracted a lot of attention recently~\cite{nandkishore_many-body_2015}.

Following Ref.~\cite{luck_investigation_2016}, we adopt here a different point of view
and consider how the system evolves as a whole, starting from a set of simple states $\left\{\ket{a}\right\}$ forming a ``preferential'' basis for the model. In the case of a particle system on a lattice, for instance, a natural choice for the preferential basis is the set of product states where the particles have fixed positions in real space. In a spin system (see below), one may choose the spin configurations
that are eigenstates of all the $S^z_i$ operators, for some choice of the quantization axis $z$.
A quantity of interest is the typical time-average probability to return to the initial basis state.
As will be explained below, these probabilities are related to
the inverse participation ratios (IPRs) of the eigenstates, computed in the preferential basis.
In the present study we explore numerically these quantities in a particular many-body problem, the spin-$\frac{1}{2}$ XXZ spin chain. The preferential basis is chosen to be the set of ``Ising configurations'' which are eigenstates of
the $z$ component of the on-site magnetization.
This spin chain Hamiltonian depends on an anisotropy parameter $\Delta$.
The main results of the paper concern the scalings of these IPRs in the gapped phase ($|\Delta|>1$) and in the gapless phase ($|\Delta|\leq1$) of the model.

The plan of this paper is the following.
Section~\ref{generalities} is a review of isolated quantum systems,
motivating the study of the IPR $t_n$ of individual energy eigenstates and of their sum $T$.
In Sec.~\ref{xxz} we recall the definition of the XXZ spin chain
and comment on the question of degeneracies.
We then present and discuss our numerical results on various observables
in the gapped phase and in the gapless phase in the main section (Sec.~\ref{num}).
Section~\ref{beyond} contains numerical results on a non-integrable spin chain
with second-neighbor interactions,
and we briefly discuss our findings in Sec.~\ref{conclusions}.

\section{Generalities}
\label{generalities}

We consider an isolated quantum system with a finite-dimensional Hilbert space of dimension $D$. The eigenstates of the Hamiltonian $\mathcal{H}$
are denoted by $\ket{n}$, and we assume for simplicity that they are non-degenerate.
In the preferential basis $\left\{\ket{a}\right\}$, the IPR of an eigenstate is by definition
\begin{equation}
t_n=\sum_{a=1}^D \left|\left< a | n \right>\right|^4.
\end{equation}
The maximum value of this quantity is reached when the eigenstate coincides with a single basis state. This is the completely localized case and gives $t_{\rm max}=1$.
On the other hand, the minimum value of $t_n$ is reached for eigenstates which are uniform superpositions of all the basis states, with the same modulus $|\langle a|n\rangle|=1/\sqrt{D}$.
This maximally delocalized limit gives $t_{\rm min}=1/D$.
These IPRs have been extensively used to measure the localization properties of a single-particle wave-function~\cite{bell_atomic_1970,edwards_numerical_1972}
(for a review in the context of the Anderson localization see, for instance, Ref.~\cite{evers_anderson_2008}).
They have also recently proved to be useful in the context of many-body localization~\cite{de_luca_ergodicity_2013,luitz_many-body_2015}.
They can be used as well to extract some universal long-distance properties from (clean) many-body ground-state wave functions~\cite{stephan_phase_2011}.

Following Ref.~\cite{luck_investigation_2016} (see also \cite{dukesz_interplay_2009}), we sum the $t_n$ over all the eigenstates to get
\begin{equation}
T=\sum_n t_n=\sum_{a,n=1}^D \left|\left< a | n \right>\right|^4 \label{eq:T}.
\end{equation}
This quantity measures how much the eigenstates are localized in the preferential basis.
It can range from $T_{\rm min}=1$ (all the eigenstates spread maximally over the whole basis)
to $T_{\rm max}=D$ (each eigenstate matches exactly a single basis state).
This quantity was originally introduced~\cite{luck_investigation_2016}
from a dynamical point of view, as the trace of the matrix $Q$ whose entries $Q_{ab}$ are the time-average probabilities
to go from state $\ket{a}$ at the initial time to state $\ket{b}$ at time $t$.
If the system is prepared in state $\ket{a}$ at $t=0$, the probability to observe it in state $\ket{b}$ at time $t$ is
$P_{ab}(t)=\left|\langle b |\psi(t)\rangle\right|^2$, {\it i.e.,}
\begin{equation}
P_{ab}(t)
=\sum_{m,n} e^{i(E_n-E_m)t} \langle b |m\rangle\langle m |a\rangle\langle a |n\rangle\langle n |b\rangle.
\end{equation}
In the absence of degeneracies in the spectrum,
the time-average probability reads
\begin{equation}
Q_{ab}=\lim_{t\to\infty} \frac{1}{t}\int_0^t P_{ab}(t')dt'
=\sum_n \left|\langle a |n\rangle\right|^2 \left|\langle b |n\rangle\right|^2.
\end{equation}
In this dynamical picture, $T/D$ measures the stationary return probability to an initial basis state, averaged over all the basis states (initial conditions).
The minimum value $T_{\rm min}/D=1/D$ is reached if the dynamics connects any initial basis state to all the other basis states, a limit of perfect ``equilibration''.
On the other hand, the maximum value $T_{\rm max}/D=1$ is reached when $\mathcal{H}$ is diagonal in the preferential basis, so that
the system does not evolve at all if prepared at $t=0$ in a basis state.

This quantity $T$ was studied in detail in the case of a single particle (tight-binding model) in a one-dimensional random potential~\cite{luck_investigation_2016}.
In the present work, we analyze the quantity $T$ for a simple {\em many-body} problem {\em without} disorder, the spin-$\frac{1}{2}$ XXZ spin chain.

\section{XXZ spin chain}
\label{xxz}

\subsection{Hamiltonian}

We consider the spin-$\frac{1}{2}$ XXZ chain with open boundary conditions, with a Hamiltonian given by
\begin{equation}
\mathcal{H}=\sum_{i=1}^{N-1} \left( S_{i}^{x}S_{i+1}^{x} +S_{i}^{y}S_{i+1}^{y} +\Delta S_{i}^{z}S_{i+1}^{z}\right).
\label{eq:HXXZ}
\end{equation}
As is well known, the system has a gapped spectrum for $|\Delta|> 1$,
with long-range order and spontaneous symmetry breaking in the thermodynamic limit (``Ising'' phase).
On the other hand, the system has a gapless spectrum and displays an algebraic decay of spin-spin correlations at zero temperature
for $|\Delta|\leq 1$ [the so-called Tomonaga-Luttinger liquid (TLL) phase~\cite{giamarchi_quantum_2004}].
Hereafter, we consider the ``Ising configurations'' $\ket{a}=\ket{\uparrow\uparrow\downarrow\cdots},\ket{\uparrow\downarrow\uparrow\cdots},\cdots$ (eigenstates of all the $S_{i}^{z}$)
as a natural basis for this problem. We use this basis to define the IPR $t_n$ of the eigenstates.

\subsection{Remarks on degeneracies}

In the case of a degenerate multiplet of eigenstates $\ket{n_1},\cdots,\ket{n_d}$,
the corresponding contribution to $T$ (as derived from the dynamical point of view~\cite{luck_investigation_2016}) should be
$
\sum_a \left(\sum_{\alpha=1}^d \left|\left< a | n_\alpha \right>\right|^2 \right)^2
$.
For open boundary conditions and a generic value of $\Delta$, it turns out that the only degeneracies
in the spectrum of Eq.~(\ref{eq:HXXZ}) are the two fold degeneracy ($S^z_{\rm tot} \leftrightarrow -S^z_{\rm tot}$) of the eigenstates with a non 	zero magnetization.
But since we focus here on the $S^z_{\rm tot}=0$ sector, we generically get non degenerate eigenstates
which are either even or odd under a global spin flip, as well as even or odd under the spatial (left-right) symmetry.

We finally note that some additional degeneracies occur at the free-fermion point ($\Delta=0$). For simplicity, we avoid this special point when numerically computing  the IPR.

\section{Numerical results}
\label{num}

\subsection{The quantity $T$}\label{ssec:T}

The eigenstates of Eq.~(\ref{eq:HXXZ}) were obtained using a full
numerical diagonalization for even sizes up to $N=20$
in the $S^z_{\rm tot}=0$ sector.
The total Hilbert space dimension is therefore
$D={{N}\choose{N/2}}\sim 2^N\,\sqrt{2/(\pi N)}$. Using the spin-flip parity and the spatial parity with respect to the center of the chain,
the total space is decomposed into four blocks, each one being labeled by two parities. 
The results for $T$ are summarized in Fig.~\ref{fig:T}.
Note that $T$ is independent of the sign of $\Delta$.
This follows
from the fact that changing the sign of $\Delta$ is equivalent to changing the sign
of the $xy$ terms, and the latter can be undone by some unitary transformation ($\pi$ rotation about the $z$ axis
on every second site) which only affects the sign of the wave functions in the Ising basis, not their modulus.
We can therefore restrict ourselves to $\Delta\geq0$.

\subsubsection{Gapped phase}

We observe (top panel of Fig.~\ref{fig:T}) that,
for sufficiently large $\Delta$, $\ln(T)$ is approximately proportional to the number of sites, {\it i.e.,}
\begin{equation}
\ln(T)\approx a(\Delta)\,N,
\label{adef}
\end{equation}
with $0<a(\Delta)<\ln(2)$.
We can deduce that the average IPR $\bar t=T/D$ behaves as $\ln(\bar t)\approx -[\ln(2)-a(\Delta)]N$.
We conjecture that these scalings hold in the whole gapped (massive) region ($\Delta>1$).

At $\Delta=\infty$, any Ising configuration (basis state) becomes an eigenstate of $\mathcal{H}$, so that one may expect $T=D$ and $\ln(T)/N\approx\ln(2)$ in that limit.
This simple reasoning is, however, not correct
since many energy levels become degenerate in this limit. Thus, as soon as $\Delta$ is not strictly infinite, the actual eigenstates
are nontrivial superpositions of several basis states, due to the effect of the $S^+_i S^-_{i+1}+{\mathrm H.c.}$ terms (according to degenerate perturbation theory).
We indeed have $\ln(T)/N\approx a(\infty)$ for very large $\Delta$,
but the numerically obtained value in this regime,
$a(\infty)\approx0.25$ (see Fig.~\ref{fig:T}),
is significantly smaller than $\ln(2)$.
A very similar phenomenon has already been observed
in the much simpler situation of a tight-binding particle
in a random potential with binary disorder~\cite{luck_investigation_2016},
where the hybridization of degenerate molecular states
has been shown to result in a nontrivial asymptotic return probability $Q\approx0.373$
in the limit of an infinitely strong disorder.

\subsubsection{Gapless phase}

\begin{figure}[h]
\includegraphics[width=\linewidth]{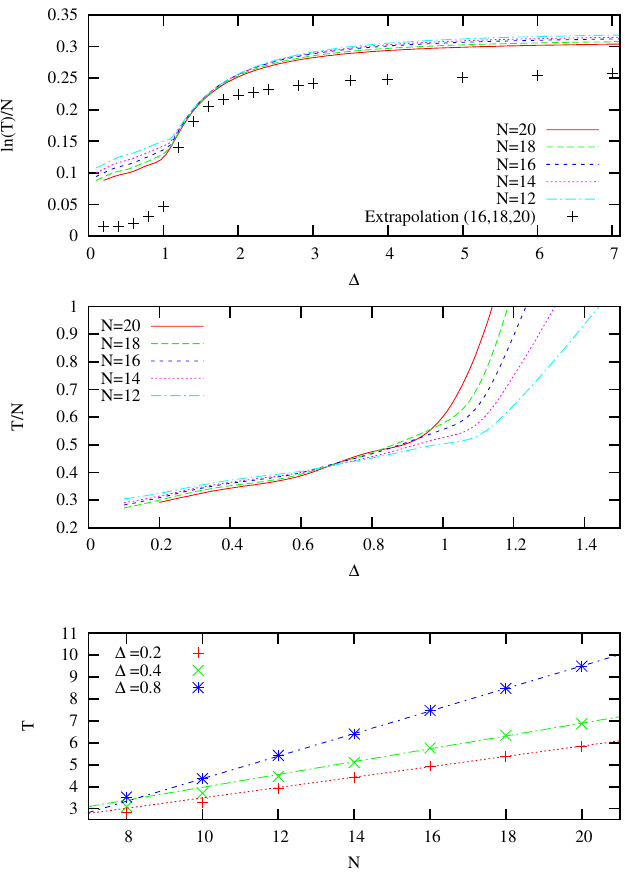}
\caption{Quantity $T$ [defined in Eq.~(\ref{eq:T})], plotted as a function of the system size $N$ (up to $N=20$ spins) and of the anisotropy parameter $\Delta$.
Different scalings are used: $\ln(T)/N$ in the top panel and $T/N$ in the middle.
In the top panel, the crosses are the results of a simple extrapolation to $N=\infty$ using $\ln(T)\approx a(\Delta)N+b(\Delta)+c(\Delta)/N$ and
three different system sizes ($N$=16, 18 and 20).
The bottom panel shows $T$ as a function of $N$ for three values of $\Delta$ in the gapless phase. The lines are guides to the eye
and show that $T$ is compatible with $T\sim N$ in this regime.
The data are computed using all the eigenstates in the $S^z_{\rm tot}=0$ sector.
}
\label{fig:T}
\end{figure}

In the gapless phase, $T$ appears to scale approximately as $N$ (bottom panel in Fig.~\ref{fig:T}).
With the present finite-size data, it is however not possible to decide whether
the exponent is exactly 1.
In any case, this implies that $a(\Delta)=0$ and that the average IPR $\bar t=T/D$ scales as $\ln(\bar t)\approx -N\ln(2)$ in the whole gapless phase.

The gapped and gapless phases are well known for having different behaviors at zero temperature (correlation functions, etc.).
This sharp distinction between $|\Delta|\leq1$ and $|\Delta|>1$, however, becomes a smooth crossover at finite temperature.
Since Eq.~(\ref{eq:T}) involves a sum over all the eigenstates, which is reminiscent of an infinite-temperature quantity,
the observation that $T$ shows qualitatively distinct behavior in both phases is unexpected and remarkable. As discussed in
Sec.~\ref{sec:entropy}, it implies a qualitative change in the distribution of the IPR of highly excited eigenstates.

\subsubsection{Remarks concerning the free-fermion point}

Finding the scaling of $T$ is a nontrivial question
even at the free-fermion point ($\Delta=0$).

Let us, however, notice that
the IPR $t_0$ of the ground state of a periodic chain of $N$ sites
(with $N$ even) at $\Delta=0$ can be computed exactly.
It is, indeed, related to the partition function of the Dyson-Gaudin gas at inverse temperature $\beta=4$~\cite{gaudin_gaz_1973},
as well as to the Shannon-Rényi entropy for the special value $n=2$ of the Rényi parameter~\cite{stephan_shannon_2009}.
The result reads
\begin{equation}
t_0=\frac{N!}{(N/2)!(2N)^{N/2}}
\end{equation}
and gives for large $N$
\begin{equation}
\ln(t_0)\approx -\frac{N}{2}+\frac{\ln(2)}{2}.
\label{eq:log_t0}
\end{equation}
We are thus facing an explicit example of an eigenstate with $\alpha=1/2$ (see below).

From another perspective,
an approximate determination of the quantity $T$ at the free-fermion point
can be obtained by replacing the two matrices defining the Bogoliubov transformation
which diagonalizes the quadratic fermionic Hamiltonian
with two independent random orthogonal matrices.
Skipping every detail, let us mention that this approach yields
$T_{{\rm SO}(N)}=\frac{1}{2}\mean{\det(1+\Omega)^2}$,
where the brackets denote an average over the orthogonal matrix $\Omega$
with the uniform (Haar) measure on ${\rm SO}(N)$.
The known value of the above average~\cite{KS} leads to the simple expression
\begin{equation}
T_{{\rm SO}(N)}=N+1.
\end{equation}
In spite of its approximate character,
this result corroborates the observed scaling $T\sim N$ in the gapless phase.

\subsection{Distribution of the IPR of individual eigenstates}
\label{ssec:tn}

\begin{figure}[h]
\includegraphics[width=\linewidth]{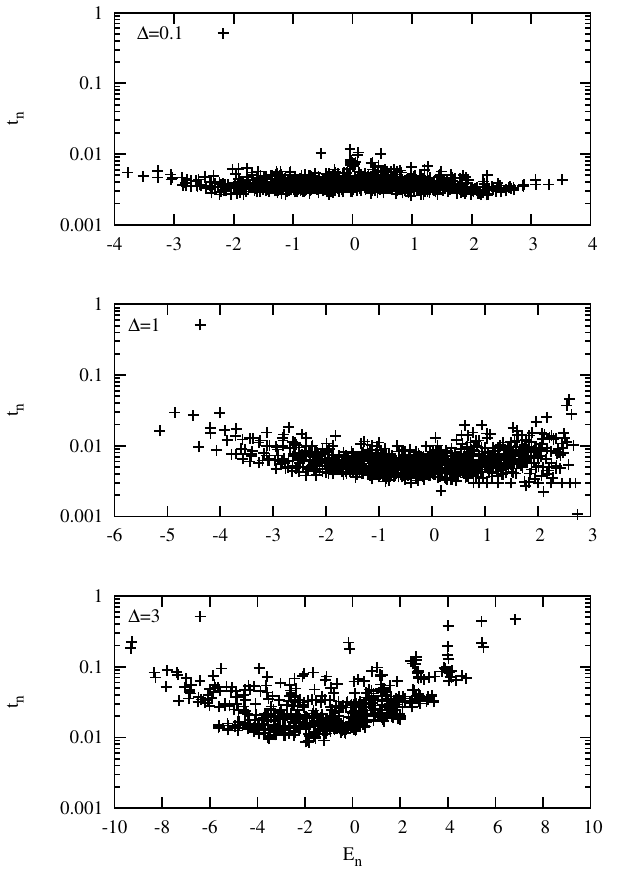}
\caption{IPR $t_n$ versus the eigenstate energy $E_n$, for all the eigenstates with $S^z_{\rm tot}=0$ in a 12-site chain.
Top to bottom: $\Delta=0.1$, 1, and 3.
The energy is weakly correlated with $t_n$ in the gapless phase.
On the other hand, for large $\Delta$, the states at the edges of the spectrum appear to be more localized (larger $t_n$) than those in the middle of the spectrum.
}\label{fig:tn_sz0}
\end{figure}

\begin{figure}[h]
\includegraphics[width=\linewidth]{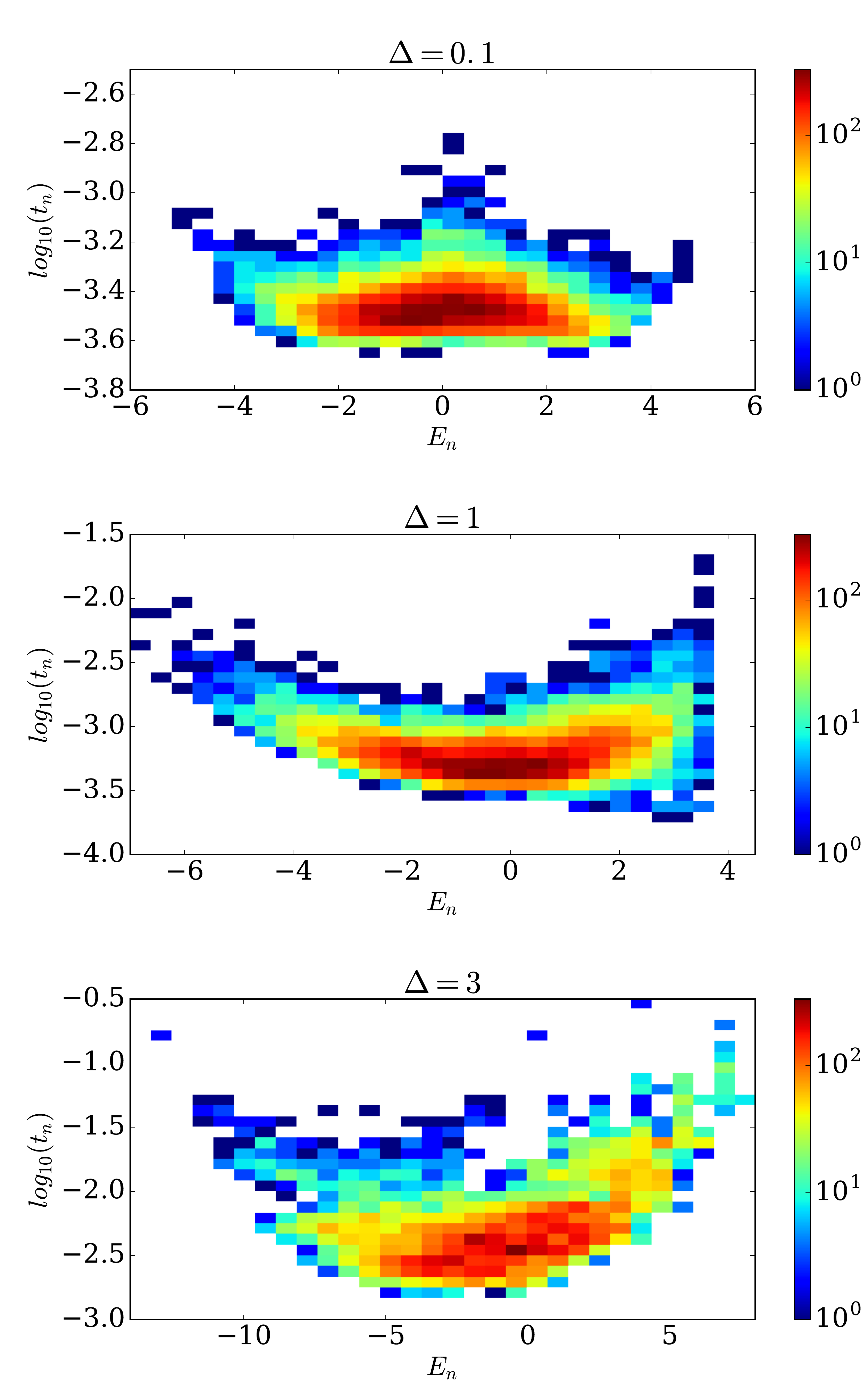}
\caption{Color plot of the density of states as a function of the IPR $t_n$ and the eigenstate energy $E_n$ for all the eigenstates with $S^z_{\rm tot}=0$ in an 18-site chain.
Top to bottom: $\Delta=0.1$, 1, and 3.
}\label{fig:tn_sz0_18}
\end{figure}

In order to understand the different scalings observed in the previous section, it is instructive to analyze how the IPRs $t_n$ are distributed over the energy spectrum.
This information is represented in Figs.~\ref{fig:tn_sz0} and \ref{fig:tn_sz0_18} for different values of $\Delta$.
Figure~\ref{fig:tn_sz0} corresponds to a 12-site chain, and each individual eigenstate is represented by a cross. Figure~\ref{fig:tn_sz0_18}
corresponds to 18 spins, and there the density of states is represented by a color scale.
For $\Delta$ close to zero it appears
that $t_n$ is weakly correlated with energy and that the IPRs of all the eigenstates are of the same order of magnitude.
For $N=16$ and $\Delta=0.1$, $t_n$ range from $0.00022$ to $0.0016$,
which corresponds to only a factor 7 .

For a generic delocalized state,
the IPR $t_n$ is expected to scale as $\exp(-\alpha_n N)$.
For instance, Eq.~(\ref{eq:log_t0}) gives an example of a state with $\alpha=1/2$.
It therefore seems natural to introduce the quantity
\begin{equation}
\alpha_n=-\ln(t_n)/N
\end{equation}
in order to compare the distributions of $t_n$ for systems with different sizes.
In the example above ($\Delta=0.1$), $\alpha$ goes from $\alpha_{\rm min}=-\ln(0.0016)/16\approx0.402$ to $\alpha_{\rm max}=-\ln(0.00022)/16\approx0.526$.

\begin{figure}[h]
\includegraphics[width=\linewidth]{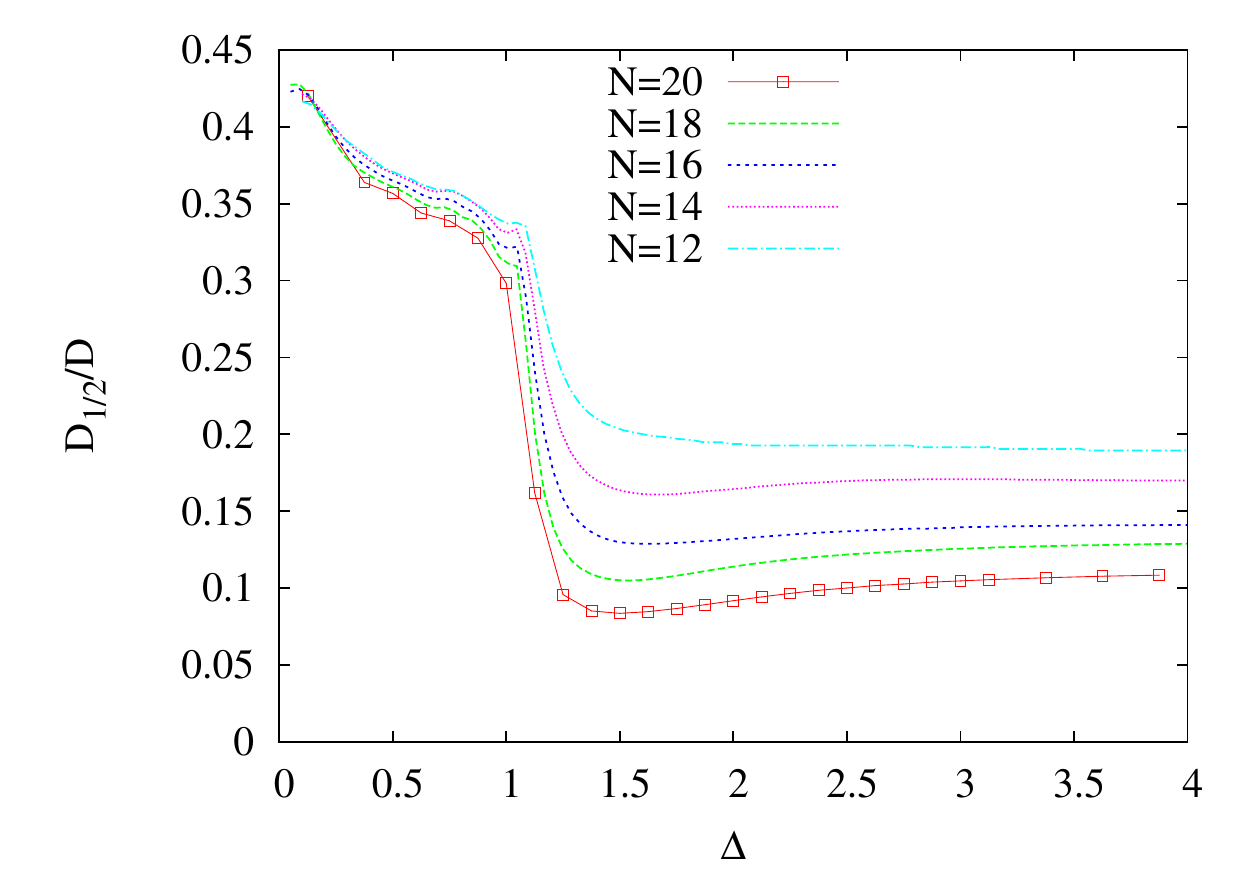}
\caption{Proportion $D_{1/2}/D$ of states needed to get $T/2$ [see Eq.~(\ref{dhalf})],
as a function of $\Delta$ and for different system sizes $N=12,14,16,18$, and 20.
The calculation is restricted to $S^z_{\rm tot}=0$.
This proportion weakly depends on $N$ for $\Delta<1$, whereas it decreases significantly with $N$ in the massive phase.
}
\label{fig:D0.5}
\end{figure}

In order to get an idea of the distribution of  $t_n$,
we propose sorting them in decreasing order and look for the number of states $D_{1/2}$
one has to include in the sum (starting from the largest $t_n$) in order to get one half of~$T$:
\begin{equation}
\sum_{n=1}^{D_{1/2}} t_n \approx \frac{T}{2}.
\label{dhalf}
\end{equation}
We find
$D_{1/2}/D\approx 0.42$ for $\Delta=0.1$ and 0.36 for $\Delta=0.5$
(see Fig.~\ref{fig:D0.5}).
Importantly, these ratios are relatively stable when varying the system size from $N=12$ to $N=20$. For this reason we conjecture that the ratio $D_{1/2}/D$ remains finite in the thermodynamic limit in the whole gapless regime.

The situation turns out to be qualitatively different for large $\Delta$ (see the bottom panels of Figs.~\ref{fig:tn_sz0} and \ref{fig:tn_sz0_18}, corresponding to $\Delta=3$). There, the eigenstates
with extremal energies (close to the ground state or to the highest excited state) are significantly more ``localized'' (larger $t_n$) than those
living in the middle of the spectrum (where the density of states is maximal).
For $N=16$ and $\Delta=3$, $t_n$ range from $0.0016$ to $0.47$,
which now corresponds to a factor 300.
For this value of $\Delta$, the proportion of states required to get one half of $T$
is $D_{1/2}/D=0.19$ for $N=12$,  0.14 for $N=16$, and 0.10 for $N=20$.
(see Fig.~\ref{fig:D0.5}). This steady decrease suggests that the ratio $D_{1/2}/D$ vanishes in
the thermodynamic limit, so that an infinitesimal proportion of all the eigenstates accounts
for the major contribution to $T$. We conjecture that this is the case in the whole massive phase.

\subsection{Entropy function of the IPR}
\label{sec:entropy}

\begin{figure}[h!]
\includegraphics[width=\linewidth]{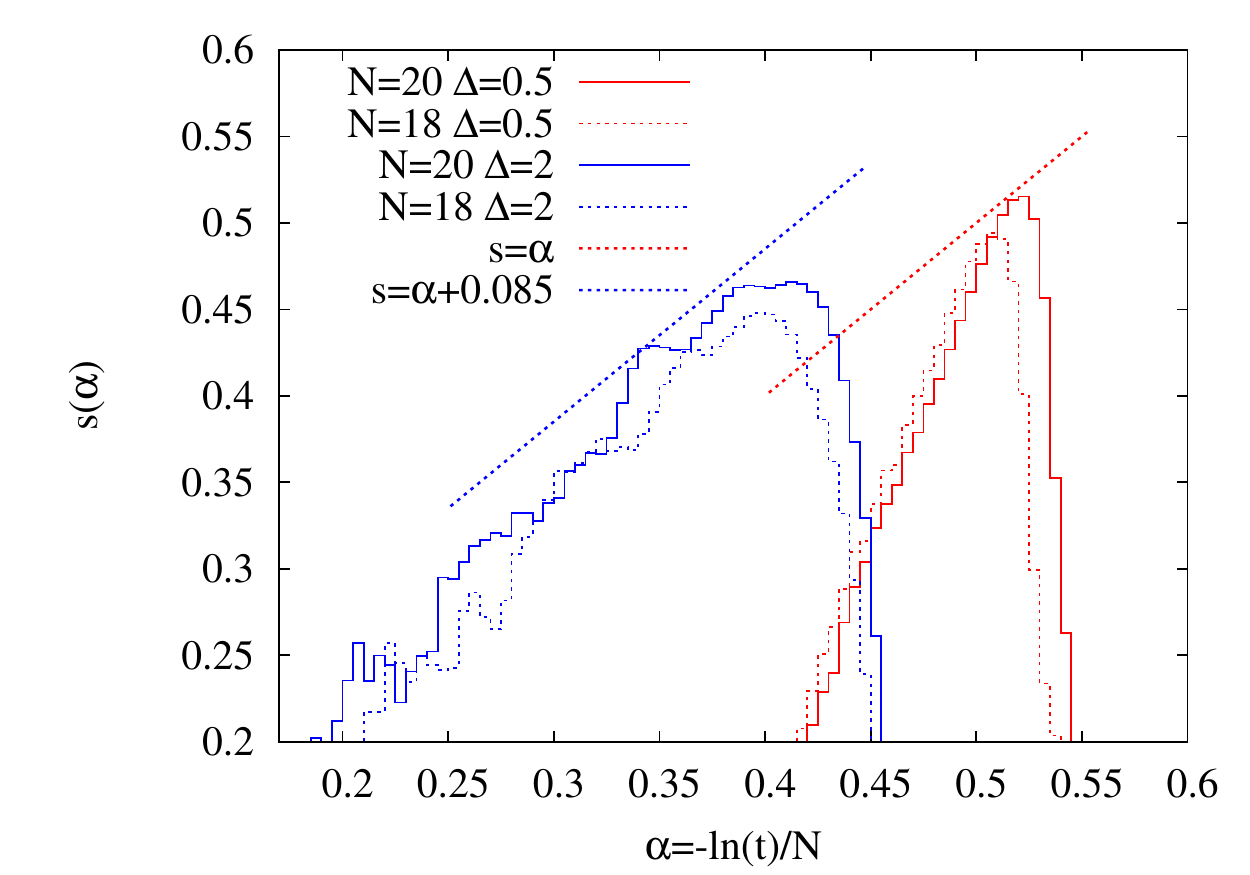}
\caption{Entropy $s(\alpha)$ [see Eq.~(\ref{sdef})] for two system sizes and two values of $\Delta$. Here the bin size is $1/200$.
The dotted lines have slope $1$ and are guides to the eye to locate the saddle point $s'(\alpha_0)=1$.
In the thermodynamic limit we expect the maximum of the entropy to converge to $s=\ln(2)$.
In addition, we conjecture that, in the thermodynamic limit and in the gapless phase,
this maximum is attained at $\alpha_0=\ln(2)$, where a cusp should develop (see text and Fig.~\ref{fig:schema}).
}
\label{fig:histo}
\end{figure}

\subsubsection{Generalities on entropy function}

The number $W(\alpha,\delta)$ of states with
a value of $\alpha$ in the interval $[\alpha,\alpha+\delta]$ is expected to scale exponentially with the system size,
in analogy with a conventional microcanonical density of states.
This leads us to define an entropy $s(\alpha)$, such that
\begin{equation}
W(\alpha,\delta)\sim e^{Ns(\alpha)}.
\label{sdef}
\end{equation}
For sufficiently large $N$, the entropy defined in this way should no longer depend on the ``bin size'' $\delta$.
As we will see, this function is the proper way to describe the distribution of $t_n$ in the thermodynamic limit.
This  entropy function can also be viewed as a large deviation function and is
similar to the quantities used in multifractal analysis.

Some numerical results for this quantity are plotted in Fig.~\ref{fig:histo}. Although we observe significant finite-size effects, the picture that emerges is the following:
\begin{itemize}
\item[(i)]
The entropy converges to a well-defined intensive function $s(\alpha)$
on an interval $[\alpha_{\rm min},\alpha_{\rm max}]$ in the thermodynamic limit.
A precise determination of this support is, however, not simple from the available data
due to the small number of eigenstates close to the extremal values of $\alpha$.
\item[(ii)]
The entropy $s(\alpha)$ has a maximum for some $\alpha=\alpha_{\rm typ}$,
with $\alpha_{\rm min}<\alpha_{\rm typ}<\alpha_{\rm max}$, where
these three quantities depend on $\Delta$.
From a saddle-point evaluation of the total number of states
\begin{equation}
D\sim\int e^{Ns(\alpha)}d\alpha\sim 2^N,
\end{equation}
we predict that the maximum $s(\alpha_{\rm typ})$ of the entropy
converges to $\ln(2)$ for sufficiently large systems.
The numerical values for $N=18$ and $N=20$ (see Fig.~\ref{fig:histo}) are smaller than $\ln(2)$, but they significantly increase with $N$.
Moreover, $\alpha_{\rm typ}$ corresponds to the typical IPR: $t_{\rm typ}\sim\exp(-\alpha_{\rm typ}N)$.
\end{itemize}

The observations made previously on $T$ can be related to properties of $s(\alpha)$.
The quantity $T$ can be estimated by evaluating the integral
\begin{equation}
T\sim\int e^{N(s(\alpha)-\alpha)}d\alpha
\end{equation}
by means of a saddle-point approximation.
The saddle point $\alpha_0$ is defined as the solution of $s'(\alpha_0)=1$
and gives $T\sim \exp\left\{N [s(\alpha_0)-\alpha_0]\right\}$.
In other words, $a(\Delta)$ [defined in Eq.~(\ref{adef})] is given by
$a(\Delta)=s(\alpha_0)-\alpha_0$.

\subsubsection{Gapless phase}

\begin{figure}[h]
\includegraphics[width=0.75\linewidth]{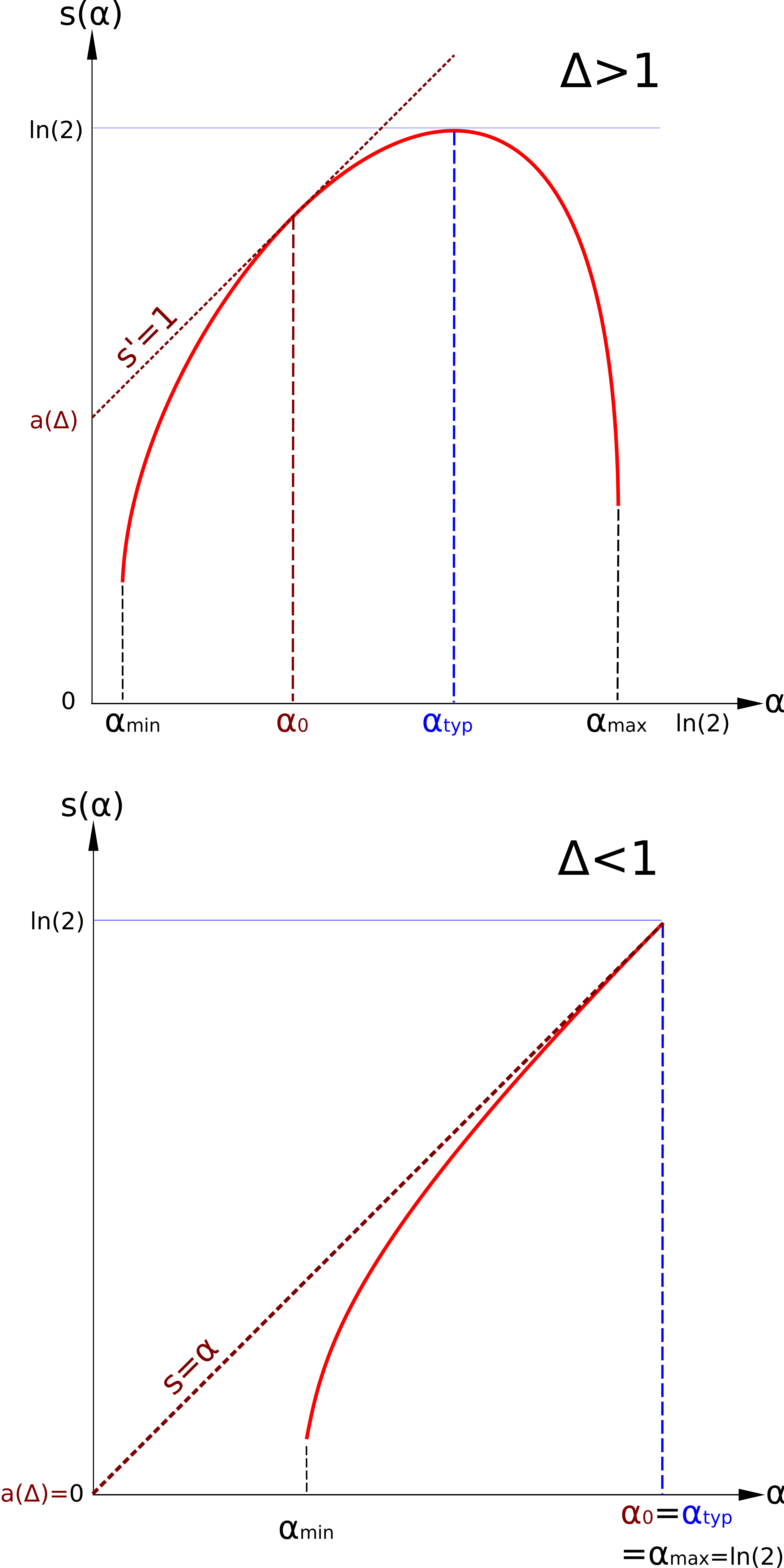}
\caption{Schematic representation of the entropy function,
as conjectured from the numerical analysis.
Top: $|\Delta|>1$ (gapped phase). Bottom: $|\Delta|<1$ (gapless phase).}
\label{fig:schema}
\end{figure}

The numerical data in Figs.~\ref{fig:T} and~\ref{fig:D0.5}
led us to conjecture that $a(\Delta)=0$ and that $D_{1/2}/D$ reaches a finite limit in the gapless phase.
These observations can be translated as follows in terms of $s(\alpha)$.
By definition, to get one half of $T$, one needs to sum over the $D_{1/2}$ states with the highest $t_n$ (lowest $\alpha_n$).
This amounts to integrating over $\alpha$ only up to some value $\bar \alpha$.
Within the saddle-point approximation,
it is easy to see that we have, in fact, $\bar\alpha=\alpha_0$.
This implies $D_{1/2}\sim \exp\left[N s(\alpha_0)\right]$.
The ratio $D_{1/2}/D$ thus scales as $\exp\{N[s(\alpha_0)-\ln(2)]\}$.
Thus, from the fact that $D_{1/2}/D$ appears to be finite in the gapless phase,
we conclude that $s(\alpha_0)=\ln(2)$.
Since $\ln(2)$ is the maximal value for the entropy function,
we also have $\alpha_0=\alpha_{\rm typ}$.
However, by definition, $s'(\alpha_0)=1$.
Thus, the entropy function has to have a cusp at its maximum.
The fact that $\alpha_{\rm typ}=\ln(2)$
also implies that the IPR of typical eigenstates
is close to that of the maximally delocalized states.
Finally, this also implies that $a(\Delta)=0$,
which is consistent with the scaling of $T$ as a power of $N$ in the gapless phase,
as already discussed.
The scenario we propose for the gapless phases is summarized
in the bottom panel of Fig.~\ref{fig:schema}.

\subsubsection{Gapped phase}

Repeating the above arguments in the gapped phase, the observed scaling for
$D_{1/2}/D$ leads to $s(\alpha_0)<\ln(2)$. The corresponding scenario
for the entropy function is summarized in the top panel of Fig.~\ref{fig:schema},
with a strictly positive value for $a(\Delta)=s(\alpha_0)-\alpha_0$
accounting for the exponential growth of $T$.
It is interesting to note that, although $T$ grows exponentially with the system size,
virtually all eigenstates have an exponentially small IPR, that is, a delocalized character.

\section{Beyond the XXZ chain}
\label{beyond}

In order to test the robustness of the results we have obtained for the XXZ chain,
we consider some perturbation of the model by including second-neighbor interactions.
The Hamiltonian is now defined as follows:
\begin{eqnarray}
\mathcal{H} &=& \sum_{i=1}^{N-1} \left( S_{i}^{x}S_{i+1}^{x} + S_{i}^{y}S_{i+1}^{y} +\Delta S_{i}^{z}S_{i+1}^{z}\right)
\nonumber\\
&+& J_2\sum_{i=1}^{N-2} \left( S_{i}^{x}S_{i+2}^{x} + S_{i}^{y}S_{i+2}^{y} +\Delta S_{i}^{z}S_{i+2}^{z}\right).
\label{eq:H_J2}
\end{eqnarray}
The parameter $J_2$ allows one to tune the strength of the perturbation.

For some detailed discussion about the rich phase diagram of this model,
we refer the reader to Refs.~\cite{hikihara_ground-state_2001,furukawa_ground-state_2012} and references therein,
but we simply focus here on the regime
$0\leq J_2\lesssim 0.24$. There, the ground state  turns out to be a gapless TLL
for $\Delta\leq 1$, while it is a gapped state (Ising phase) for $\Delta>1$. In other words, for sufficiently small $J_2$, the
zero-temperature phase diagram is similar to the $J_2=0$ case. Note, however, that the model is not integrable for $J_2\ne 0$.

In the TLL phase it is conventional to
parametrize  the long-distance properties of the correlation functions  by the so-called Luttinger parameter $K$.
Using a bosonization approach, one can show that the staggered part of $\mean{S^+_0S^-_r}$ then
behaves as $\sim  (-1)^r r^{-\frac{1}{2K}}$, and that of $\mean{S^z_0S^z_r}$ behaves
as $\sim  (-1)^r r^{-2K}$~\cite{giamarchi_quantum_2004}. $K$ is {\it a priori} some 
nontrivial function of the microscopic parameters of the model, here $\Delta$ and $J_2$. However, the
SU$(2)$ symmetry present at $\Delta=1$ forces the correlations to be isotropic, so that, in fact, $2K=\frac{1}{2K}$ and $K=\frac{1}{2}$.
Next, as discussed by Haldane~\cite{haldane_spontaneous_1982}, the  perturbation (umklapp terms) that drives the transition from the TLL to the Ising
phase is  marginal (in the renormalization-group sense) when $K=\frac{1}{2}$. For this reason,
the transition line between the TLL and the Ising phase lies exactly at $\Delta=1$.

\begin{figure*}[!htbp]
\includegraphics[width=\linewidth]{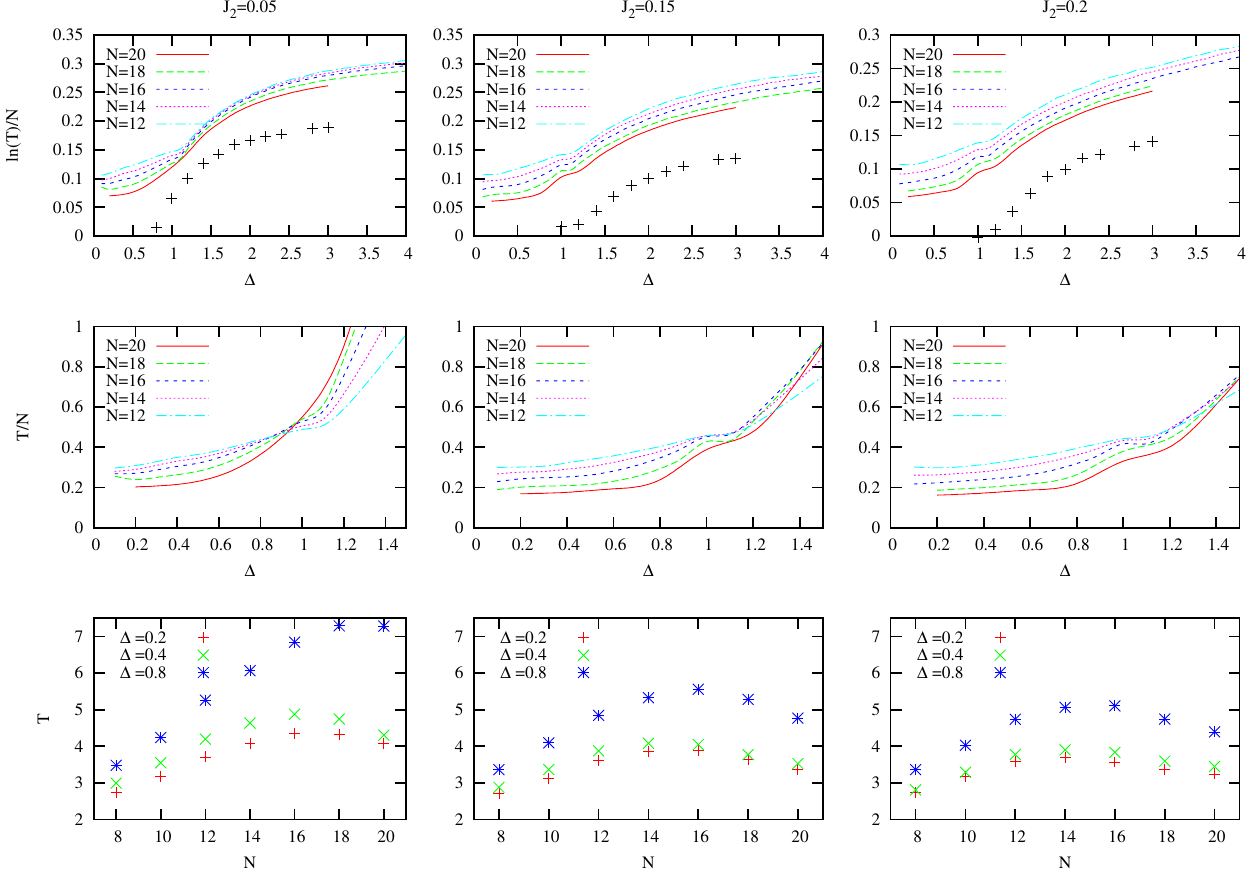}
\caption{Same as Fig.~\ref{fig:T} for a spin chain with second-neighbor interactions
[see Eq.~(\ref{eq:H_J2})].
From left to right: $J_2=0.05, 0.15$, and $0.2$. The crosses in the top panels represent some extrapolations
based on $N=16,18$, and $20$, as in Fig.~\ref{fig:T}.
}
\label{fig:T_J2}
\end{figure*}

The numerical results concerning the scaling of $T$ for $J_2=0.05, 0.15$, and $0.2$ are displayed
in Fig.~\ref{fig:T_J2}. We first note that the ``perturbed'' model displays an exponential growth of $T$ [$T\sim \exp(a N)$ with $a>0$] in the gapped phase, as for
the $J_2=0$ case. This is rather clear from the plots in the top panels and from the associated extrapolation points (crosses).

The situation, however, seems different for $\Delta<1$.
The bottom panels of Fig.~\ref{fig:T_J2} indeed show that $T$ reaches a maximum and then slightly decreases for larger system sizes.
Since $T$ cannot vanish (it is larger than $1$ by construction), these data strongly suggest that $T$ is finite
in the thermodynamic limit when $J_2>0$.
How can this be reconciled with the observation that $T$ grows linearly with $N$ when $J_2=0$ and $|\Delta|<1$ (Fig.~\ref{fig:T})~?
The first possibility is that the $N\to\infty$ value of $T$ is finite for the non-integrable models but diverges as $J_2$ approaches zero.
This interesting scenario would imply a strong effect of the integrability of the model on the IPR distribution.
We note that the integrability has been shown in Ref.~\cite{santos_onset_2010} to have an impact on participation ratios, even though that
study focused on different basis choices.

An alternative, albeit less probable, possibility is that $T$ also remains finite in the gapless phase of the integrable model ($J_2=0$).
Although this is not what the data of Fig.~\ref{fig:T} (bottom panel) suggest, one cannot exclude that $T$ converges to a finite limit for larger systems.
If that were the case, studying larger chains would be needed to observe some saturation of $T$ for the unperturbed XXZ chain.
Still, performing a full diagonalization on open chains beyond $N=20$ sites would be numerically quite challenging.
Indeed, for 22 spins at $S^z_{\rm tot}=0$, the Hilbert space dimension
is larger than $1.7\times10^5$ in each of the four symmetry sectors (using space inversion and global spin flip).

\section{Conclusions}
\label{conclusions}

We have investigated the IPRs (inverse participation ratios)
of individual energy eigenstates in a preferential Ising basis
for a spin-1/2 XXZ chain without disorder
by means of an exact diagonalization of the Hamiltonian on finite chains of length $N$ (up to 20),
with open boundary conditions.
We have considered in particular the sum $T$ of all $t_n$,
which has a dynamical interpretation:
it yields the stationary return probability to a typical initial state
of the preferential basis.
Our main finding is the observation of a qualitatively different behavior
of the latter quantities in the gapped and gapless phases.
In the gapped phase ($|\Delta|>1$),
$T$ grows exponentially with $N$, and the entropy function $s(\alpha)$
describing the distribution of $t_n$ has a smooth maximum.
In the gapless phase ($|\Delta|<1$),
$T$ seems to scale linearly with $N$,
whereas the entropy function is singular at its maximum.
We have also investigated the effect of next-nearest-neighbor interactions,
which break the integrability of the model.
Although $T$ still grows exponentially in the gapped phase,
it now appears to saturate to a constant value in the gapless phase.

In some future work it would be very useful to
make some progress concerning the free-fermion point ($\Delta=0$)
as well as in the limit $\Delta\to\infty$. As we explained, as far as the ground-state
wave function(s) are concerned, these two limits are rather simple. However, computing $T$ for these models
appears quite challenging.
More generally, we need some deeper understanding of the qualitatively different scalings
observed in the gapped and gapless phases. From this point of view, a Bethe ansatz formulation of the eigenstates
seems a promising route to explore, as well as some continuum-limit approach to the problem in the gapless phase.
The latter may also shed light on the possible existence of universal terms in the quantity $T$.

\bibliography{trQ.bib}{}

\end{document}